\documentclass[preprint,10pt]{aastex}
\pdfoutput=1

\received{}
\revised{}
\accepted{}

\slugcomment{Accepted for publication in the Astrophysical Journal}

\shortauthors{Wood et al.}
\shorttitle{Photoionization Simulations of a SN Driven Turbulent ISM}

\begin{document}

\title{Photoionization of High Altitude Gas in a 
Supernova Driven Turbulent Interstellar Medium}

\author{Kenneth Wood\altaffilmark{1,2},  Alex S. Hill\altaffilmark{2}, M. Ryan Joung\altaffilmark{3, 4}, 
Mordecai-Mark Mac Low\altaffilmark{4}, Robert A. Benjamin\altaffilmark{5}, L. Matthew Haffner\altaffilmark{2}, 
R.J. Reynolds\altaffilmark{2}, G.J. Madsen\altaffilmark{6}
}

\altaffiltext{1}{School of Physics \& Astronomy, University of St Andrews, 
North Haugh, St Andrews, Fife, KY16 9AD, UK}

\altaffiltext{2}{Department of Astronomy, University of Wisconsin-Madison, 475 N. Charter St, 
Madison, WI 53706}

\altaffiltext{3}{Department of Astronomy, Columbia University, 550 West 120th Street, 
New York, NY 10027, USA}

\altaffiltext{4}{Department of Astrophysics, American Museum of Natural History, 
79th Street at Central Park West, New York, NY 10024, USA}

\altaffiltext{5}{Department of Physics, University of Wisconsin-Whitewater, Whitewater, WI 53190}

\altaffiltext{6}{Sydney Institute for Astronomy, School of Physics, University of Sydney, NSW 2006, Australia}

\email{kw25@st-andrews.ac.uk}

\begin{abstract}

We investigate models for the photoionization of the widespread diffuse ionized gas in galaxies. 
In particular we address the long standing question of the 
penetration of Lyman continuum photons from sources close to the galactic midplane 
to large heights in the galactic halo. We find that recent hydrodynamical simulations of a supernova-driven 
interstellar medium have low density paths and voids that allow for ionizing 
photons from midplane OB stars to reach and ionize gas many kiloparsecs above the midplane. 
We find ionizing fluxes throughout our 
simulation grids are larger than predicted by one dimensional slab models, thus allowing for photoionization 
by O stars of low altitude neutral clouds in the Galaxy that are also detected in H$\alpha$. In previous 
studies of such clouds the photoionization scenario had been rejected and the H$\alpha$ had been attributed 
to enhanced cosmic ray ionization or scattered light from midplane H~{\sc ii} regions. 
We do find that the emission measure distributions in our simulations 
are wider than those derived from 
H$\alpha$ observations in the Milky Way. 
In addition, the horizontally averaged height dependence of the gas density in the hydrodynamical 
models is lower than inferred in the Galaxy. These discrepancies are likely due to the absence of magnetic 
fields in the hydrodynamic simulations and we discuss how magnetohydrodynamic effects may reconcile 
models and observations. Nevertheless, we anticipate that the inclusion of magnetic fields in the dynamical 
simulations will not alter our primary finding that midplane OB stars are capable of producing high altitude 
diffuse ionized gas in a realistic three-dimensional interstellar medium.

\end{abstract}


\maketitle

\section{Introduction}

Diffuse ionized gas (DIG, also commonly referred to as the Warm Ionized Medium in the Milky Way) 
is observed along all lines of sight in the Milky Way \citep{hrt03}. Observations of 
H$\alpha$ emission \citep{hrt99}, pulsar dispersion measures \citep{r89a,gbc01,gmc08,sw09}, and [\ion{Al}{3}] absorption \citep{sw09} indicate the DIG in the Galaxy has a scale height of $1-1.5$~kpc and an H$^+$ column density about $1/3$ that of the neutral hydrogen \citep{r91a}. 
Widespread DIG is a common feature in spiral galaxies. It is detected to large heights in edge-on galaxies 
\citep{Rand98} and in the inter-arm regions of face-on galaxies well away from traditional \ion{H}{2} regions 
\citep{Ferguson96}. 
For a summary of the physical properties (ionization state, electron temperature, scaleheight) of the DIG in the 
Milky Way and other galaxies see the recent review by \citet{hdb09}. 

The power requirements of the DIG indicate that the only reasonable source of ionization is from O stars 
\citep{r90b,r90a}. However, in a smoothly distributed interstellar medium (ISM) with a hydrogen number density of 
1~cm$^{-3}$, an O star with an ionizing luminosity $Q=5\times 10^{48}\,{\rm s}^{-1}$ can only ionize a 
volume of radius 53~pc (assuming a hydrogen recombination coefficient 
$\alpha_B = 2.59 \times 10^{-13}\,{\rm cm}^3\,{\rm s}^{-1}$). 
Photoionization models including O stars in a 
vertically stratified ISM \citep[e.\ g.][]{mc93,ds94,wm04,cbf02} show that ionizing photons can reach 
to large distances perpendicular to the midplane, but in general require a lower H{\sc i} density than inferred 
from observations. Clumping in a three dimensional (3D) ISM naturally provides low density paths for ionizing 
photons to reach large heights above the midplane. Indeed, models that adopt an average ISM density 
re-arranged into a 3D fractal structure readily allow photons to penetrate from the midplane to large heights \citep[e.\ g.\ ][]{cbf02,hdb09}. Turbulence is perhaps the most plausible mechanism for producing such clumping and the DIG is well known to be turbulent \citep{ars95,b99,hbk08,cl10}. The concept of 3D structures allowing 
deeper penetration of radiation compared to smooth media is not new or unique to photoionization studies. 
Previous theoretical work has demonstrated the increased mean free paths for non-ionizing photons in 
3D dust structures in the ISM  and molecular clouds \citep{Boisse1990}, stellar winds \citep{Shaviv2000}, 
and exoplanetary atmospheres \citep{Hood2008}. 

In this paper we extend our investigation of O star photoionization modeling of a 3D ISM to study 
photoionization of 3D hydrodynamic simulations of supernova-driven turbulent ISM \citep{jm06,jmb09}. 
In \S~2 we briefly describe the hydrodynamical and photoionization simulations; \S~3 presents the results of 
our photoionization simulations showing the resulting distributions of ionized and neutral gas; in \S~4 we 
discuss discrepancies between our models and observations of the Galactic DIG and the likely role of 
magnetic fields not present in the current simulations.

\section{Models}

\subsection{Hydrodynamic Simulations}

In this paper we investigate the ionization structure of a 3D density field from a hydrodynamic simulation of a supernova-driven turbulent ISM \citep{jm06,jmb09}. The simulations use the FLASH v2.4 code \citep{for00}. 
We summarize the essential features of the models here. 
The adaptive mesh simulation has a maximum resolution of just under 2 pc in a region with $|z| < 250$~pc, 
and progressively coarser resolution at larger altitudes. The full-resolution simulation box 
has dimensions 0.5~kpc in $x$ and $y$ and extends to $\pm$5~kpc in $z$ above and below the midplane. 
A fixed gravitational potential from \citet{kg89} was used. The model employed here, with the Galactic 
supernova rate, has Type~II and Type~I supernova rates of $27.4$ and 
$6.58 \textrm{ Myr}^{-1} \textrm{ kpc}^{-2}$ and scale heights of $90$ and $325$~pc, respectively. 
Of the Type~II supernovae, $3/5$ are concentrated spatially and temporally to simulate superbubbles and 
their heights are confined to $|z|\le 90$~pc. 
The total gas surface mass density is $1.87 \times 10^6 M_{\odot} \textrm{ kpc}^{-2}.$

A diffuse heating term representing photoelectric heating \citep{Wolfire95} 
of $\Gamma = 8.50 \times 10^{-26} \textrm{ erg s}^{-1}$ and radiative cooling appropriate for an optically thin, solar metallicity plasma are included. The hydrodynamical simulation does not track ionization of the gas, so photoionization heating is not explicitly included. In the DIG, photoionization by a dilute radiation field is likely the dominant heating mechanism, although there may be an additional mechanism that is important at low densities, 
$n\la 0.1\,{\rm cm}^{-3}$
 \citep{rht99,wm04}. However, this has a small impact on the dynamics of the ISM because the thermal pressure is small compared to the turbulent pressure. In addition, \citet{sb07} have shown that magnetic fields may provide 
 additional heating in the DIG. This is very relevant especially given our conclusions on the need for the 
 inclusion of magnetic fields in the dynamical simulations to explain the observed emission measure 
 distributions in the Galactic DIG.

In their hydrodynamic simulations of a vertically stratified ISM driven by
individual supernova explosions, \citet{jm06} find that the density
power spectrum peaks at $\sim 20$ pc and that the most energy-containing scales
lie at 20-40 pc (see their Fig. 8).  On the other hand, they find no
single effective driving scale;  energy injection occurs over a broad
range of scales, with greater than 90\% of the total kinetic energy contained in
wavelengths shortward of 200 pc.  Spatially correlated supernovae explosions that
break out of the gaseous disk (i.e., superbubble blow outs) and
high-altitude Type Ia supernovae are the main drivers of turbulent motions of the
halo gas.  These structures have larger characteristic scales (several
hundred parsecs) than individual supernova remnants in the disk, due to the higher
amount of energy and lower background gas density, respectively; see Fig. 2 of \citet{jm06}.

For our photoionization simulations we consider a subset of this grid 
extending to $\pm$2~kpc in $z$. Memory requirements of our photoionization code on current desktop 
computers required us to re-bin the density grid by a factor of four giving a resolution of 7.8~pc per grid cell. The computationally straightforward approach of running the radiative transfer code in a static realization of the hydrodynamical code is justified by the recombination time scale of $\sim 3$~Myr, much shorter than the dynamical timescale of $\sim 100$~Myr in an 8000~K hydrogen gas with a total number density of $1 \textrm{ cm}^{-3}$ and $n_e = 0.03 \textrm{ cm}^{-3}$ \citep{s78}. 

Density, velocity, and temperature slices of the simulation are shown in \cite{jmb09}. 
In Fig.~1 we show the horizontally averaged density as a function of height compared to a model for the hydrogen 
density in the ISM. The model comprises a Dickey-Lockman \citep{DL90} distribution for neutral hydrogen 
with a low density, vertically extended distribution for ionized hydrogen as follows:
\begin{equation}
n(z) = 0.4\exp{\left[ -\frac{1}{2} \left( \frac{z}{90} \right)^2 \right]}+0.11\exp{ \left[-\frac{1}{2} \left( \frac{z}{225} \right)^2 \right]} + 0.06 \exp{ \left[ -\frac{|z|}{400} \right] }
+0.025\exp{\left[-\frac{|z|}{1000} \right]}\; ,
\end{equation}
where the number density is in units of cm$^{-3}$ and the heights, $z$ are in pc. The first three terms represent 
the density of neutral hydrogen and the fourth term represents ionized hydrogen. 
The mean vertical column density (within the range $|z| \le 2\,{\rm kpc}$) for the simulations is in good agreement with that derived from equation~1 with a value of 
$\sim 7\times 10^{20}\,{\rm cm}^{-2}$. However, 
the density structure of the dynamical simulation is more centrally peaked than the Galactic model described 
by equation~1. As we 
discuss later, this is most likely due to the absence of magnetic fields in the current dynamical simulations.

\subsection{Photoionization Simulations}

The primary focus of this paper is the ionization structure of hydrogen and the penetration of Lyman 
continuum photons to large heights above the Galactic midplane. Therefore for the photoionization simulations 
we use the 3D hydrogen-only Monte Carlo photoionization code described in \citet{wl00}. 
Because neutral hydrogen is opaque to Lyman continuum radiation whereas the optical depth of nearly 
fully ionized hydrogen is much lower, we use an iterative process to establish a self-consistent distribution 
of neutral and ionized hydrogen. 
This code follows the propagation of ionizing photons within a 3D linear cartesian density grid and 
accurately computes the 3D ionization structure of 
hydrogen for an assumed uniform temperature. The code includes photoionization arising from both direct 
stellar photons and also the diffuse radiation field produced by recombinations to the ground level of hydrogen. 
We have extended the original \citet{wl00} code to include the effects of dust scattering and 
absorption of ionizing photons. We adopt an opacity of 
$1.3\times 10^{-22}\, {\rm cm}^{-2}\, {\rm H}^{-1}$ appropriate for silicate dust at Lyman continuum wavelengths 
as discussed in \citet{MW05}, and assume albedo and scattering phase function 
asymmetry parameters $a=0.5$, $g=0.5$. While dust opacity may influence the ionization 
structure of high density \ion{H}{2} regions, its effects in the low density DIG that we are investigating are 
not important. We find that typically less than 1\% of ionizing photons are absorbed by dust in our 
simulations. The simulated emission measure distributions in \S~3 below do not include regions of the 
grid with $|z|< 150$~pc, thereby further mitigating the effects of dust in our analysis. 
For our current study we are not investigating emission line ratios, so the single temperature hydrogen-only code 
is sufficient. It has the advantage of being faster and 
requiring less memory than our code that computes the ionization structure of many more elements and also 
the electron temperature \citep{wme04}.

The sources of ionizing radiation are point sources within the simulation grid representing single stars or clusters 
of OB stars. We follow the approach of \citet{cbf02} and randomly choose $x$ and $y$ locations 
of the sources uniformly within the grid. The $z$ coordinate of each source is randomly chosen from a Gaussian 
distribution with a scale height of 63~pc \citep{m01}. 

The catalogue of 424 O stars within 2.5~kpc of 
the Sun \citep{gcc82} indicates a stellar surface density of 24 stars kpc$^{-2}$. The Lyman continuum 
luminosity from these stars was estimated to be $Q = 7\times 10^{51}\, {\rm s}^{-1}$ \citep{vgs96}. 
The density field from the hydrodynamic simulation represents a slice of the ISM with a 
horizontal width of 500~pc, so 
adopting the estimated stellar surface density, we would expect around six sources within our grid 
providing an ionizing luminosity of around $10^{50}\, {\rm s}^{-1}$. A fraction of these ionizing photons 
will produce the DIG, with the rest producing local H~{\sc ii} regions or escaping from the simulation entirely. 
We do not consider the formation of high density H~{\sc ii} regions around sources, so our models are really 
considering the ionizing photons that have escaped from their local environment into the lower density 
diffuse ISM. 
In our simulations we therefore select random locations for six ionizing sources. We uniformly 
distribute the ionizing luminosity among the six sources and investigate total luminosities (i.e., 
ionizing photons escaping from traditional H~{\sc ii} regions) in the 
range $0.1\le Q_{49}\le 10$, where $Q_{49} = Q/(10^{49}\, {\rm s}^{-1})$, 
covering a range of scenarios allowing for 
a lower and higher Lyman continuum budget than estimated by \citet{vgs96}.

Repeating boundary conditions are adopted 
for the Monte Carlo radiation transfer whereby ionizing photons that exit an $x$ or $y$ face of the grid are 
re-injected on the opposite face of the grid with their direction of travel unaltered. This simulates the more 
realistic situation within the Galaxy of ionizing photons that originate from distant sources. Therefore 
ionizing photons can only exit through the top or bottom $z$ faces of the grid or if they are degraded to a 
non-ionizing photon or absorbed by dust.

\section{Results}

We investigated a large number of models for different source locations and total ionizing luminosities. 
The position of the sources within the density grid and the total ionizing luminosity affect the vertical 
extent to which gas in the hydro simulation can be ionized. For some realizations of the randomly 
sampled source locations, sources may be placed many tens of pc above the midplane, and 
thus above the densest regions in the hydrodynamic simulation. Such a positioning of sources results 
in more of the gas at large heights being ionized compared to simulations where the sources are confined 
to midplane regions. Similar results were found in the ionization simulations of \citet{cbf02} who 
investigated the escape of ionizing photons from vertically stratified smooth and 3D (fractal) ISM density distributions. 

The single parameter that most strongly influenced our results and determined the vertical extent to which the 
gas could be ionized was the total ionizing luminosity, $Q_{49}$. The number and location of ionizing sources 
did not have a big effect -- if a source was randomly 
placed within a very dense region of the grid and its ionizing photons were trapped, there were always 
other sources in less dense regions that could provide the photons to ionize gas at large heights. In the 
following sections we quantitatively present the results of our photoionization simulations.

\subsection{Distribution of neutral and ionized hydrogen}

Figure~2 shows the average neutral fraction ($n[{\rm H}^0] / n[{\rm H}^0 + {\rm H}^+]$) 
as a function of height for different ionizing luminosities. 
At low values of $Q_{49}\la 1$, the grid becomes fully neutral for $|z|\ga800$~pc. 
We find that the grid is almost fully ionized at large heights ($|z| \gtrsim 0.5$~kpc) for total ionizing 
luminosities $Q_{49}\ga 2$. Figure~3 displays the emission 
measure ($\textrm{EM} \equiv \int n_e^2 ds$) and \ion{H}{1} column density maps for 
edge-on viewing of the grid, showing the development of extended ionized 
volumes as the ionizing luminosity increases.

To maintain the ionization of the Galactic DIG in the Solar neighborhood (within 2-3~kpc) 
requires a Lyman continuum flux in the midplane of around 
$4\times 10^6\,{\rm cm}^{-2}\,{\rm s}^{-1}$ \citep{r90b}, about 12\% of that available from OB stars \citep{a82,vgs96}. The required ionizing luminosity, both for the Galactic DIG and for our simulations, 
is lower than that estimated to be available from OB stars in the Galaxy and represents Lyman continuum 
photons that leak out of traditional (high density) H~{\sc ii} regions to ionize the DIG. 
As discussed in \S~2.2, assuming the \citet{vgs96} value for the ionizing flux in the 
solar neighborhood ($3.4\times 10^7\,{\rm cm}^{-2}\,{\rm s}^{-1}$), the total flux available to ionize 
our grid is around $10^{50}\,{\rm s}^{-1}$. In our simulations the gas is ionized at large heights when 
$Q_{49}\ga 2$, slightly higher than the 12\% suggested by  \citet{r90b}. 

Figure~4 shows the average neutral and ionized hydrogen densities as a function of height for the 
simulation with $Q_{49}=2$. 
As described above, the mean densities at large $|z|$ are lower than 
the model of equation~1 for hydrogen density in the Galaxy. 
Therefore it is not surprising that the mean density of 
neutral hydrogen for $|z|\ga 200$~pc in the ionization simulations is lower than the Dickey-Lockman 
distribution. The mean electron density from the simulations is also lower than derived for the DIG in 
the Galaxy, but does follow an exponential distribution at large $|z|$ with a scale height of around 500~pc. 
This is smaller than the scale height of the DIG in our Galaxy.

For comparison, we investigated ionization of a smooth density grid with the 
hydrogen density described by equation~1. 
In the simulations we adopted an ionizing luminosity $Q_{49}= 2$ and multiplied 
the first three terms in equation~1 by a constant factor until we achieved ionization of the gas at large heights 
and a total escape fraction of ionizing photons of around 10\% through the top and bottom of our grid 
($z=\pm 2$~kpc). To achieve this we find that the three components representing the ``neutral hydrogen" 
in equation~1 must be multiplied by about 1/3. This is consistent with the ionization study of 
\citet{mc93} who found that if they used the known locations of O stars in the solar neighborhood 
then the hydrogen density within the smooth component must be lower than the Dickey-Lockman 
distribution to allow gas at large heights to be ionized. Also see the results of 3D simulations by 
\citet{cbf02} and the discussion in \citet{hdb09}. For O stars to provide the ionization of 
the DIG requires a 3D distribution of gas in which the density of any smooth component should be at 
most one third of the Dickey-Lockman distribution for neutral hydrogen. To make up the remainder of the 
observed neutral hydrogen requires higher density ``clouds" \citep{mc93} or denser regions such as produced in a turbulent medium.

We further investigated the survival of neutral clouds at large heights above the midplane and performed 
simulations to study the ionization of a 50~pc radius spherical blob positioned at $z=1.5$~kpc in the grid. 
Figure~5 shows a slice through the center of the ionization grid 
containing a blob of various densities showing the transition from almost fully neutral to fully ionized with 
decreasing blob density. 
In our simulations with $Q_{49}=2$, a blob with $n\ga 0.1\,{\rm cm}^{-3}$ remains neutral, 
except for a very thin ionized skin, not resolvable in our current simulations. 
Lower density blobs exhibit a thicker ionized skin and become neutral deeper into the blob. 
The ionized skin is mostly directed towards the midplane, but does extend all around the blob due to photons 
from the diffuse ionizing radiation field. 
The number density of neutral hydrogen blobs in our simulations agrees well with density estimates 
($0.05 \la n \la 0.5$) for 
high velocity clouds \citep[e.\ g.][]{wv97,wyw08,hhr09}. 

\subsection{Emission measure distribution}

The results of the previous section demonstrate that the hydrodynamical simulations produce density 
structures that readily allow for ionizing photons to percolate to high altitudes and produce extended 
layers of ionized gas. In Figure~6 we present probability distributions of emission measure, 
${\rm EM} = \int n_{\rm e}^2\, {\rm d}s$, 
for face-on and edge-on views of the simulation grid. In forming these EM distributions we have removed 
the contributions from grid cells in the simulation with $|z| <150$~pc, which contain the densest regions 
of the simulation and in reality should also include high density \ion{H}{2} regions. In forming the EM from 
the simulations we use the computed electron density and not an EM derived from simulated H$\alpha$ 
observations. An H$\alpha$ derived EM would require an accurate temperature calculation incorporating 
heating and cooling from a radiation-magnetohydrodynamical simulation, beyond the scope of the 
current paper.

For comparison we also show EM distributions in the Galaxy derived from subsets of the Wisconsin H-Alpha Mapper (WHAM) Northern Sky Survey \citep{hrt03}.
The two EM~$\sin |b|$ distributions displayed are of high-latitude ($|b| > 10^\circ$) sightlines that 
sample 1) all DIG gas north of $\delta = -30^\circ$ and 2) the Perseus Arm, which is conveniently separated 
out in the velocity range $-100 \textrm{ km s}^{-1} \lesssim V_{\rm lsr}\le -25 \textrm{ km s}^{-1}$ \citep{hrt99}. 
In both cases, we have excluded the large classical \ion{H}{2} regions identified by \citet{hbk08} to leave a sample consisting primarily of DIG emission.

Figure~6 shows that the simulations produce EM distributions that are broader than observed 
for the Galactic DIG. In making these comparisons we compare the face-on view of the simulation with the 
EM distribution for all Galactic sightlines $b> 10^\circ$, and comparing the edge-on view with the EM distribution 
for the Perseus Arm. The very extended low EM tail seen in the edge-on view of the simulation grid will be 
diminished somewhat in H$\alpha$ derived EMs since hot ($T > 10^5$~K), low density 
($n < 10^{-2}\, {\rm cm}^{-3}$) gas at high latitudes \citep[see][]{jmb09} will have a very low 
H$\alpha$ emissivity.

The standard deviation of $\log \mathrm{EM}$ in the face-on view of the photoionization simulation is $0.56$. This is much wider than the observed value of $\sigma_{\mathrm{EM} \sin |b|} = 0.190$ for the DIG \citep{hbk08}. Indeed, in the isothermal MHD models presented by \citet{klb07} and \citet{hbk08}, only the highly supersonic models produce such a wide distribution of emission measures. Turbulence in the DIG is likely mildly supersonic, with a Mach number of $1-3$ \citep{pjn97,hbk08,bf08}. In the hydrodynamical models presented here, the warm gas, defined as $5000 \textrm{ K} < T < 20000 \textrm{ K}$, has a mean sonic Mach number of $M_s = 1.2-1.7$, as shown in Figure~\ref{fig:ms_hists}.

The breadth of the EM distributions indicates that the hydrodynamical 
simulations show much larger contrasts between low and high density gas than present in the 
Galaxy. This result, combined with the average hydrogen density at large heights 
being lower than derived in the Galaxy 
(Fig.~1), leads us to speculate that the effects of magnetic fields (not present in the simulation) could 
reconcile the model with observations. Including magnetic fields in the hydrodynamical simulations will 
provide two main effects: a higher signal speed, and thus a lower effective Mach number, which 
will result in less extreme density variations \citep[as seen in the simulations by][]{dAB05} and additional pressure support that will levitate and maintain a higher density with height. We anticipate that these combined effects will result in narrower EM distributions and higher densities with height, bringing our simulations closer to observationally derived values for the EM and height dependence of the density of the neutral and ionized gas.

\section{Comparison to previous work}

The escape fractions of ionizing photons from our simulations are around 10\% for 
$Q= 2\times 10^{49}\, {\rm s}^{-1}$ and increase for larger ionizing luminosities, entirely consistent 
with the study by \citet{cbf02}. However, ours and other 3D simulations differ from the 1D model of 
\citet{bm99} that predicted very small ionizing fluxes at small distances above the 
galactic midplane (see Fig.~6 in \citealt{pbv03}). This difference is not surprising for several reasons: 
the \citet{bm99} model considered all ionizing sources to be located in the Galactic 
midplane and to suffer extinction due to a uniform density slab of dust. This is entirely reasonable for their 
study of ionization of high velocity clouds at very large distances from the Galaxy, and for HVCs 
their model predicts appropriate escape fractions and ionizing fluxes. 
However, for studying photoionization of clouds at 
small distances from the Galactic plane their model predicts a vanishingly small flux, which is clearly not 
correct for a 3D simulation. Considering these differences between 1D and 3D models, we believe 
it is inappropriate to reject 
photoionization of the Galactic cirrus as a source for at least some of the observed H$\alpha$ emission \citep{mjl07}. Indeed, the observed H$\alpha$ intensities of typically 2~Rayleighs 
from the clouds studied by \citet{mjl07} require ionizing fluxes of around 
$5\times 10^6\,{\rm photons}\,{\rm cm}^{-2}\,{\rm s}^{-1}$, well within the range suggested for the Solar neighborhood 
by \citet{vgs96}. For a slab cloud with a uniform density of $10\, {\rm cm}^{-3}$ normally illuminated on 
one side and assuming an ionized gas temperature of 8000~K, the ionized skin would be 
$\sim 0.05$~pc thick and contribute to the observed H$\alpha$ intensity in addition to any scattered light 
component. Allowing for a photoionized skin on the surface of these clouds means that the observed 
H$\alpha$ emission need not be attributed entirely to cosmic rays 
\citep{dBC06} or scattered light from midplane H~{\sc ii} regions \citep{mjl07}.

Recent models of the DIG in M51 \citep{Seon09} suggested that O stars were not capable of producing 
ionization at large distances from their immediate surroundings. This conclusion was based on an 
ionization model of a smoothly distributed ISM (uniform slab or an exponential disk) 
that appeared to show that unrealistically small optical depths of neutral hydrogen in the ISM 
were required to allow ionizing photons to reach gas at large distances from the O stars. 
However, the analysis by \citet{Seon09} 
did not comprise a correct photoionization simulation where the ionization state of the gas was 
calculated self-consistently. Instead the penetration of ionizing photons through a neutral ISM was 
characterized using an $\exp(-\tau)$ factor where $\tau$ is the optical depth due to neutral 
hydrogen. \citet{Seon09} correctly stated that the optical depths of neutral hydrogen in a smooth ISM 
are large and will not allow photons to reach large distances from the midplane. 
Instead, \citet{Seon09} claims that optical depths 
of neutral hydrogen need to be smaller by factors of around $10^{-5}$ to allow ionizing photons to 
penetrate from midplane O stars to high altitudes in M51. 
However, in a photoionization simulation the 
neutral hydrogen fraction in diffuse ionized gas is in fact very low, typically $10^{-3}$ and even smaller 
closer to ionizing sources \citep{wm04}. Therefore 
when the medium is ionized 
Lyman continuum photons can indeed penetrate to large distances. In addition, as demonstrated in 
ours and other simulations clumping of the gas in the ISM produces low density regions and this 
further helps the ionizing photons 
reach and ionize gas at large distances from the midplane. The combination of small neutral fractions in the DIG 
and low density regions in a 3D ISM will combine to produce the low column 
densities of neutral hydrogen (compared to a smoothly distributed ISM) 
required for ionizing radiation from O stars to reach and ionize the DIG in M51.

\section{Conclusions and future work}

Our main conclusion is that the density structures produced by 3D hydrodynamical simulations of a turbulent 
ISM allow ionizing photons from midplane sources to reach and ionize gas at large altitudes. This is in 
agreement with  previous photoionization simulations of analytically produced fractal density structures that 
also indicate that a 3D ISM is required for O stars to ionize high altitude gas. We therefore believe 
that the results of the photoionization simulations of analytic ISM densities and now those of more 
realistic hydrodynamical simulations lend further compelling support for O stars being the dominant ionization 
source of the DIG.

While a 3D ISM can solve the problem of propagating ionizing photons to large altitudes, our results 
indicate that the dynamical simulations likely require additional physics to reproduce the 
details of the DIG observations within the Galaxy. Compared to the Galactic DIG, the simulations in this paper 
show broader EM distributions suggesting that magnetic fields may be the missing ingredient to reconcile 
observations with theory. Magnetic fields will reduce the density variations and provide pressure to support 
gas and give a higher density at kiloparsec heights above the midplane. This is borne out in the ISM MHD 
simulations of \citet{dAB05} and preliminary simulations of our own that incorporate magnetic fields 
(Hill et al. in preparation).

\acknowledgements

The hydrodynamical simulations in this paper use the FLASH v2.4 code developed by the Center for 
Thermonuclear Flashes at the University of Chicago \citep{for00}. 
MRJ and M-MML are partly supported by NASA/SAO grant TM0-11008X from the Chandra Theory Program. 
The Wisconsin H-Alpha Mapper, ASH, LMH, and RJR are supported by the National Science Foundation 
through grant AST-06-07512. GJM is supported by a Research Fellowship from the University of Sydney

{\em Facilities:} \facility{WHAM}


\bibliography{ms.bbl}

\begin{figure}
\includegraphics[scale=0.5,angle=90]{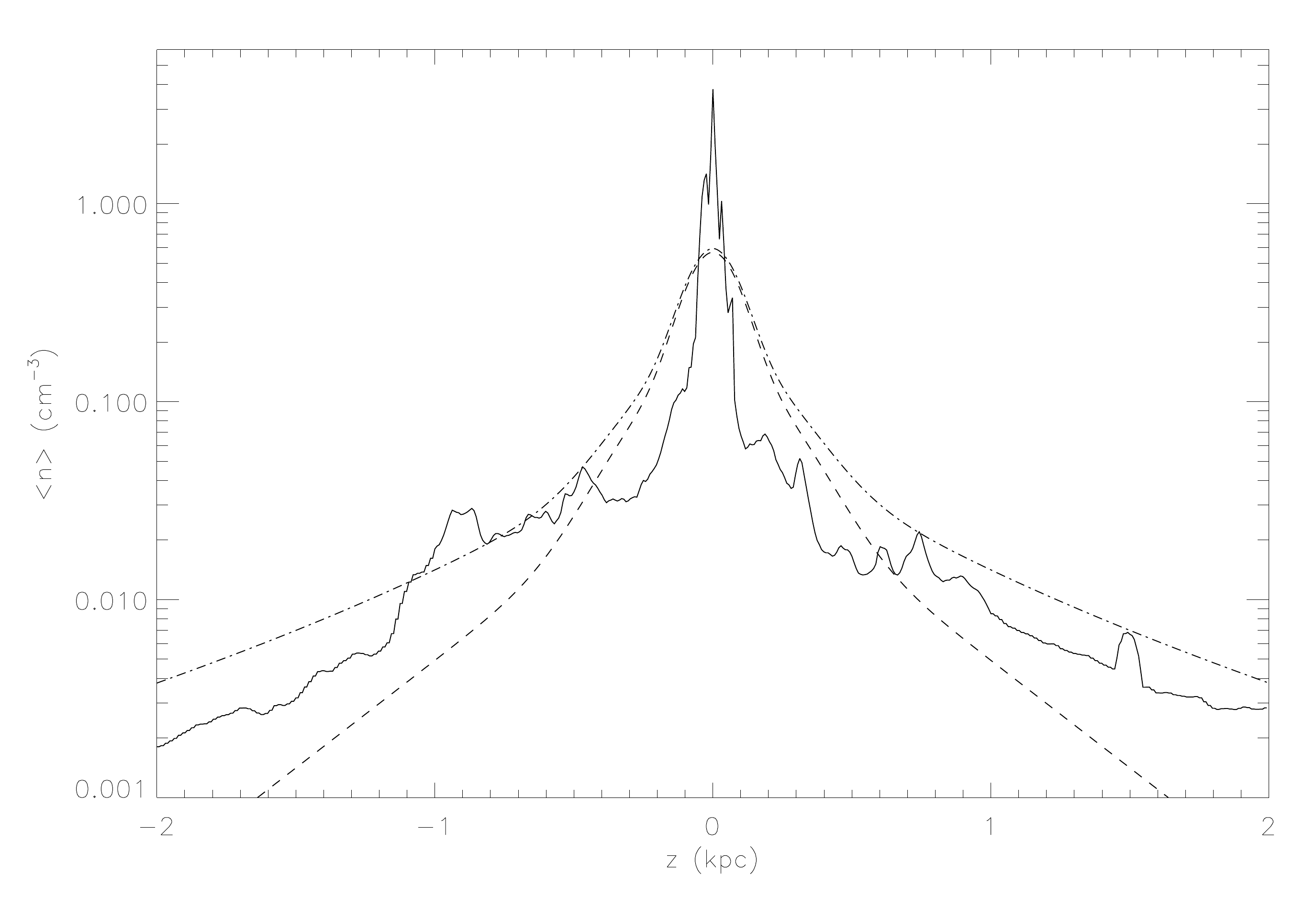}
\caption{Mean density (solid line) from the hydrodynamic simulations compared with the Dickey-Lockman 
distribution for neutral hydrogen (dashed line) and a Dickey-Lockman plus extended component representing 
diffuse ionized gas (dot-dash line).}
\end{figure}

\begin{figure}
\includegraphics[scale=0.5,angle=90]{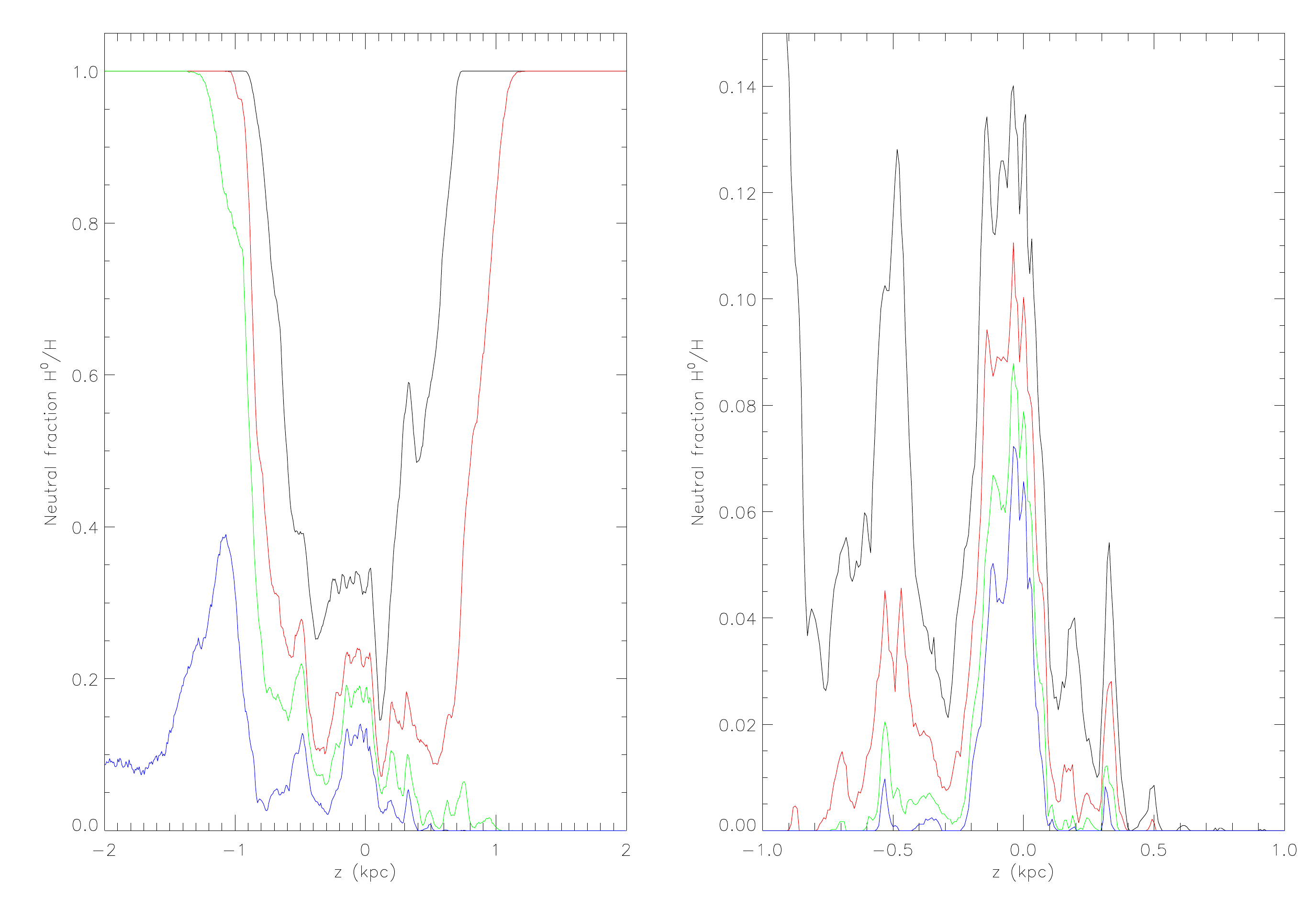}
\caption{Average neutral fraction as a function of height for ionizing luminosities (left panel)
$Q_{49}=0.1$ (black), 0.5 (red), 1 (green), 2 (blue) and (right panel) $Q_{49}=2$ (black), 3 (red), 
4 (green), 5 (blue).  }
\end{figure}

\begin{figure}
\includegraphics{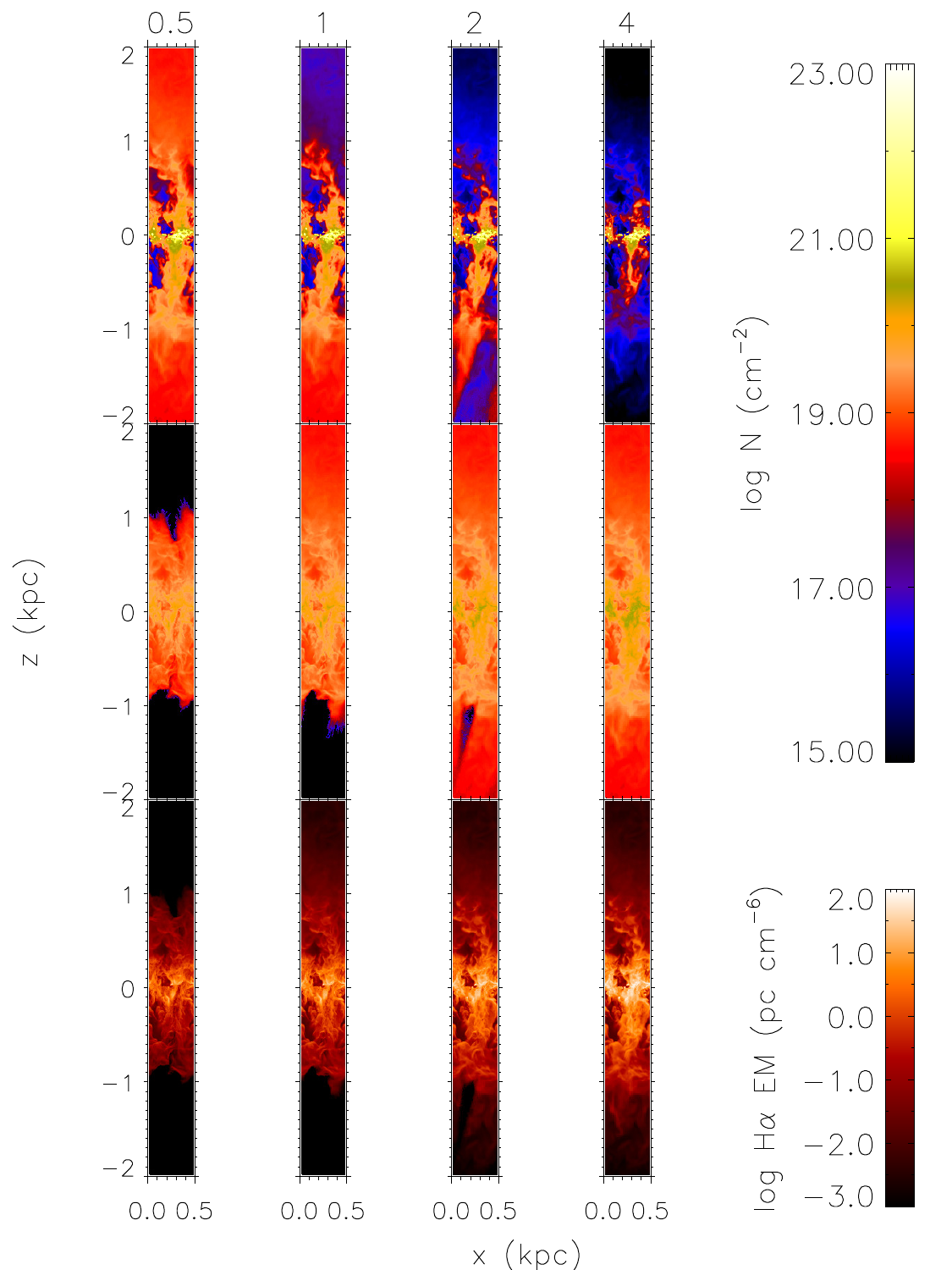}
\caption{Edge-on visualizations of the simulation grids. Column densities of neutral hydrogen (top row) and ionized hydrogen (middle row), and 
emission measure ($\int n_e^2 ds$) (bottom 
row) are shown for ionizing luminosities of (left to right) $Q_{49}=0.5, 1, 2, 4$. }
\end{figure}

\begin{figure}
\includegraphics[scale=0.5,angle=90]{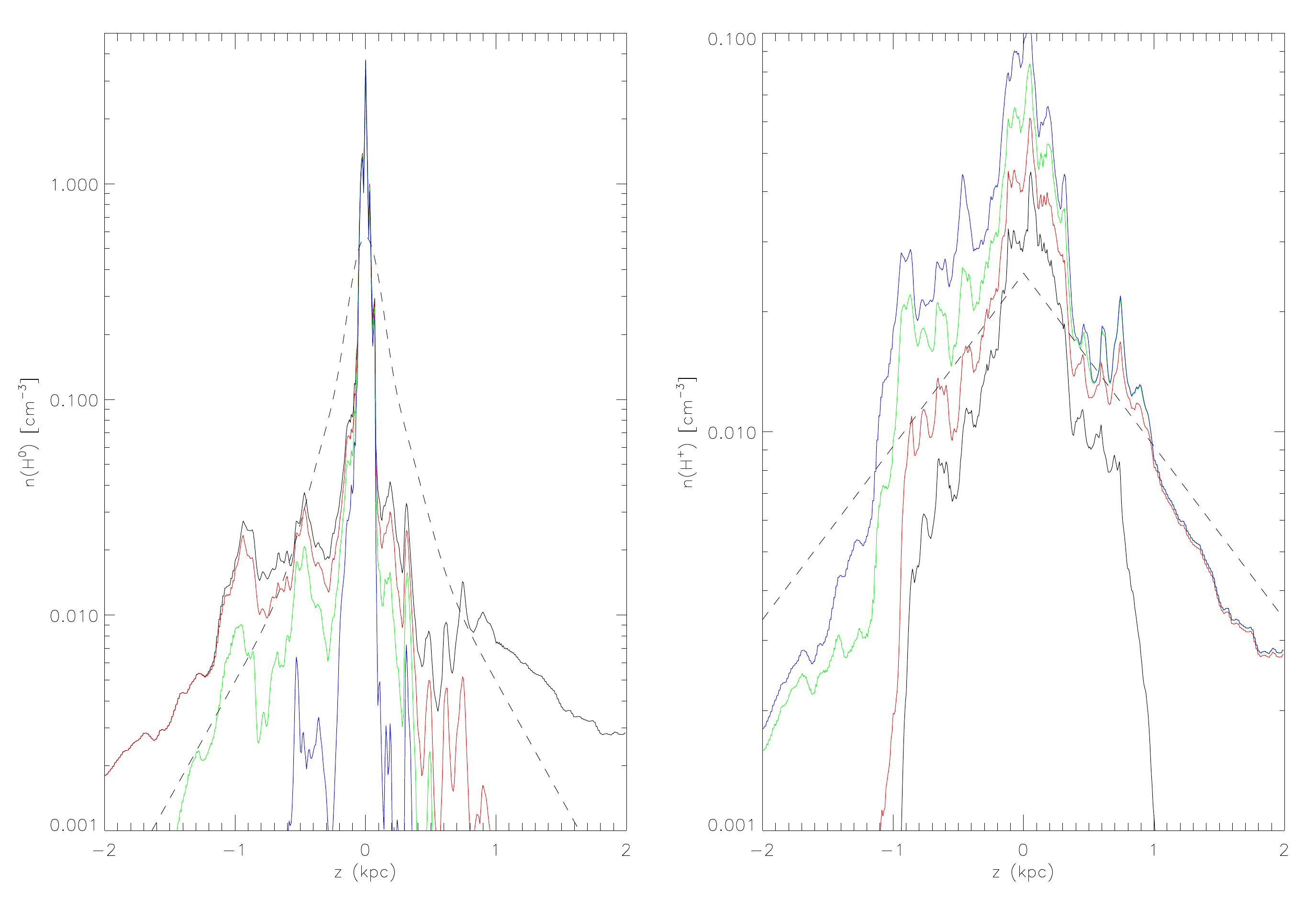}
\caption{Mean densities of neutral (left panel) and ionized hydrogen (right panel) 
for ionizing luminosities  $Q_{49}=0.5$ (black), 1 (red), 2 (green), 4 (blue). 
Overplotted with dashed lines are the Dickey-Lockman distribution for neutral hydrogen (left panel) 
and the average electron density derived for the Galactic DIG (right panel). }
\end{figure}

\begin{figure}
\includegraphics[scale=0.5,angle=90]{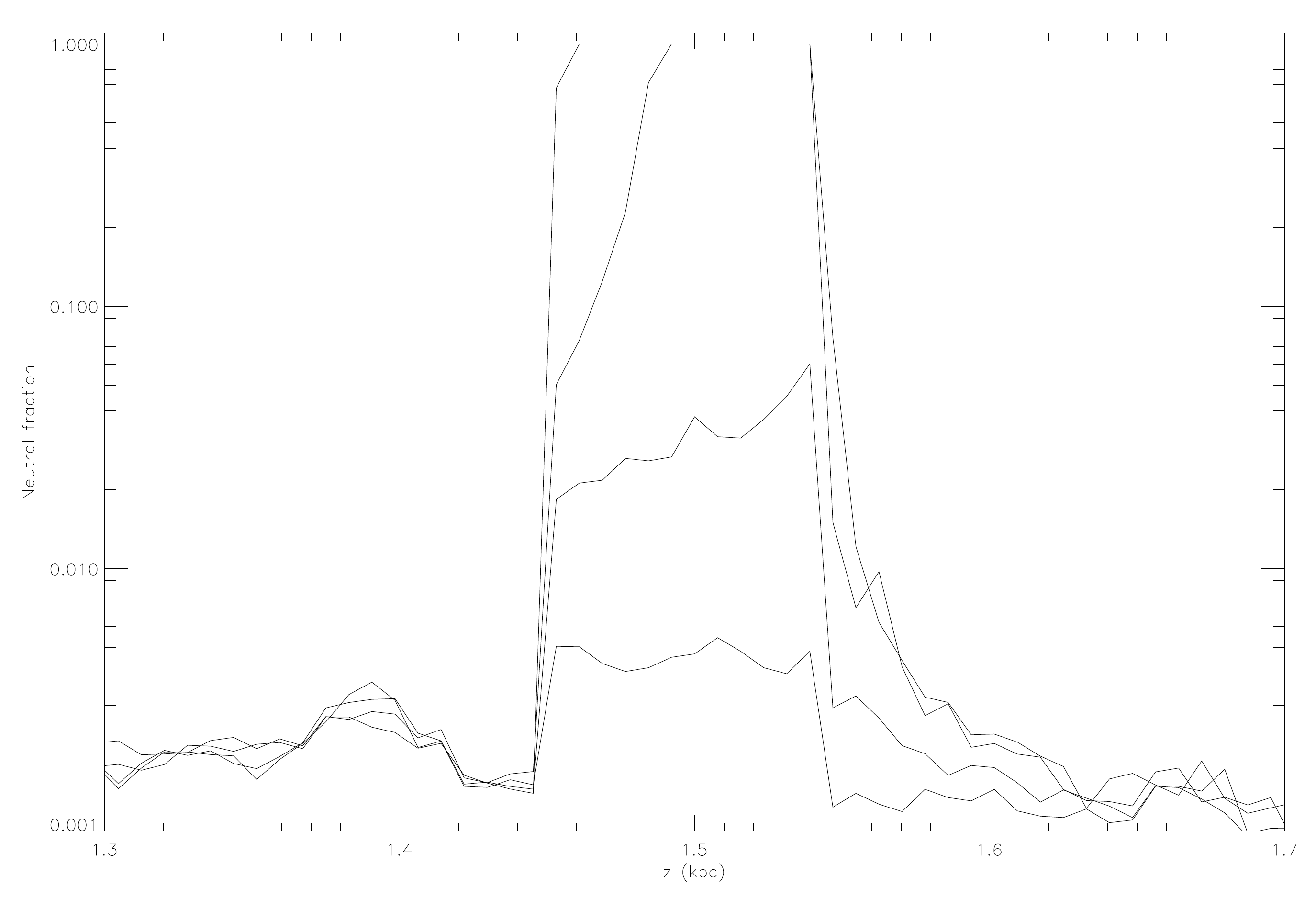}
\caption{Neutral fraction as a function of height through the center of the grid 
for simulations with a 50~pc spherical blob located at 
1.5~kpc above the midplane. The ionizing luminosity is $Q_{49}=2\,{\rm s}^{-1}$ and the 
curves (lowest to highest) are for blob densities of 0.01, 0.05, 0.1, 0.5~cm$^{-3}$. As the density 
increases the blob transitions from being almost fully ionized, to having an ionized surface skin, to being 
almost fully neutral.}
\end{figure}

\begin{figure}
\includegraphics[scale=1.0]{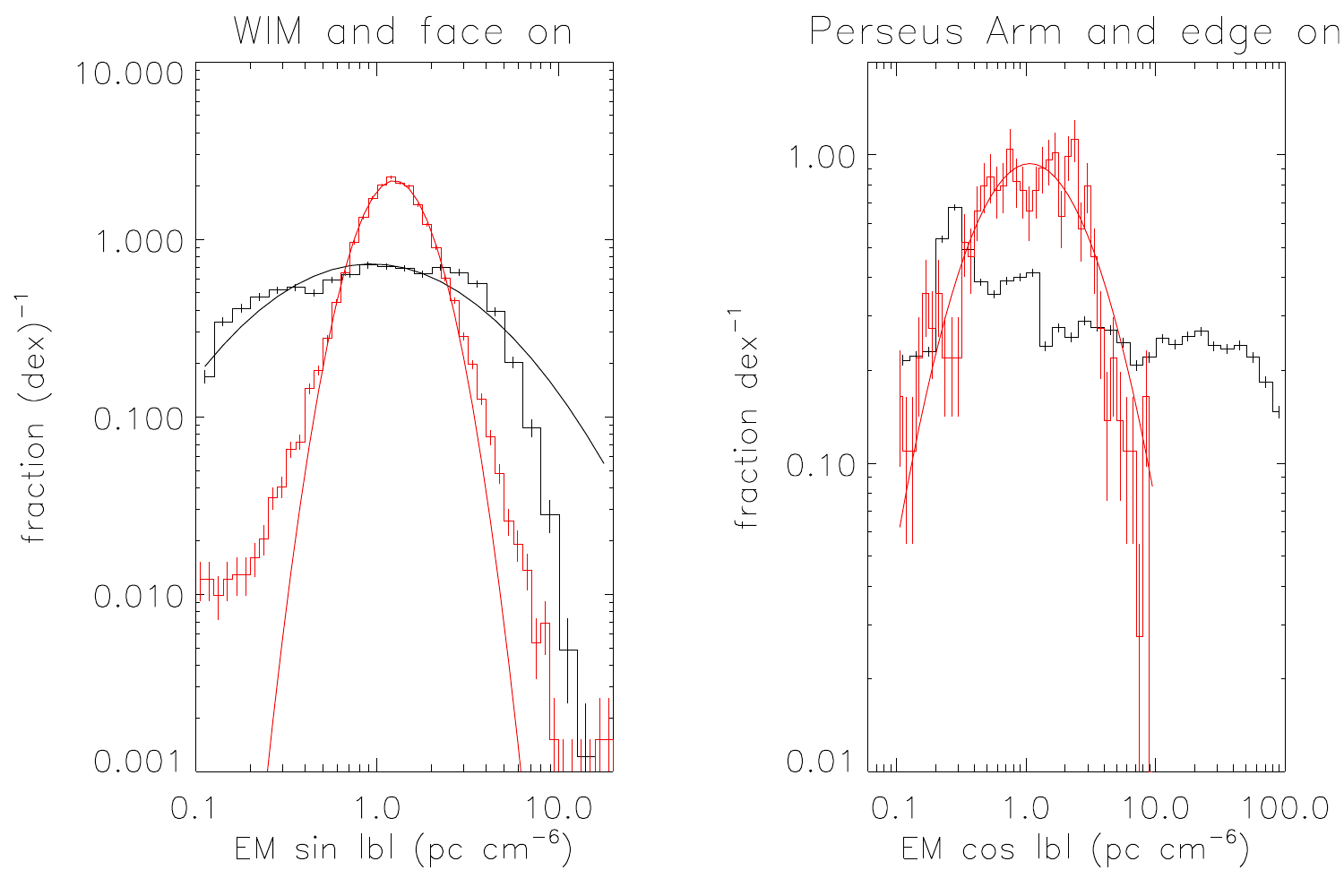}
\caption{Emission measure distributions for the Galactic DIG from the WHAM survey and from our 
$Q_{49}=2$ photoionization simulation. Fraction (dex)$^{-1}$ is the fraction of points in each bin divided 
by the logarithmic bin interval. {\em Left panel:} The black histogram is of $\int n_e^2 ds$ in the simulation, viewed face-on. The red histogram is of EM~$\sin |b|$ from the WHAM survey with classical \ion{H}{2} regions and sightlines with $|b| < 10^\circ$ removed, from \citet{hbk08}. Lognormal fits to each distribution are also shown. {\em Right panel:} The black histogram is of $\int n_e^2 ds$ in the simulation, viewed edge-on. The red histogram is of EM~$\cos |b|$ from the WHAM survey toward the Persus Arm for velocities in the range $-100 \textrm{ km s}^{-1} \lesssim V_{\rm lsr}\le -25 \textrm{ km s}^{-1}$ with classical \ion{H}{2} regions removed. For the emission measures from the simulations we have removed contributions from midplane regions with $|z|< 150$~pc. Notice that the model distributions are broader than the WHAM observations.}
\end{figure}

\begin{figure}
\includegraphics{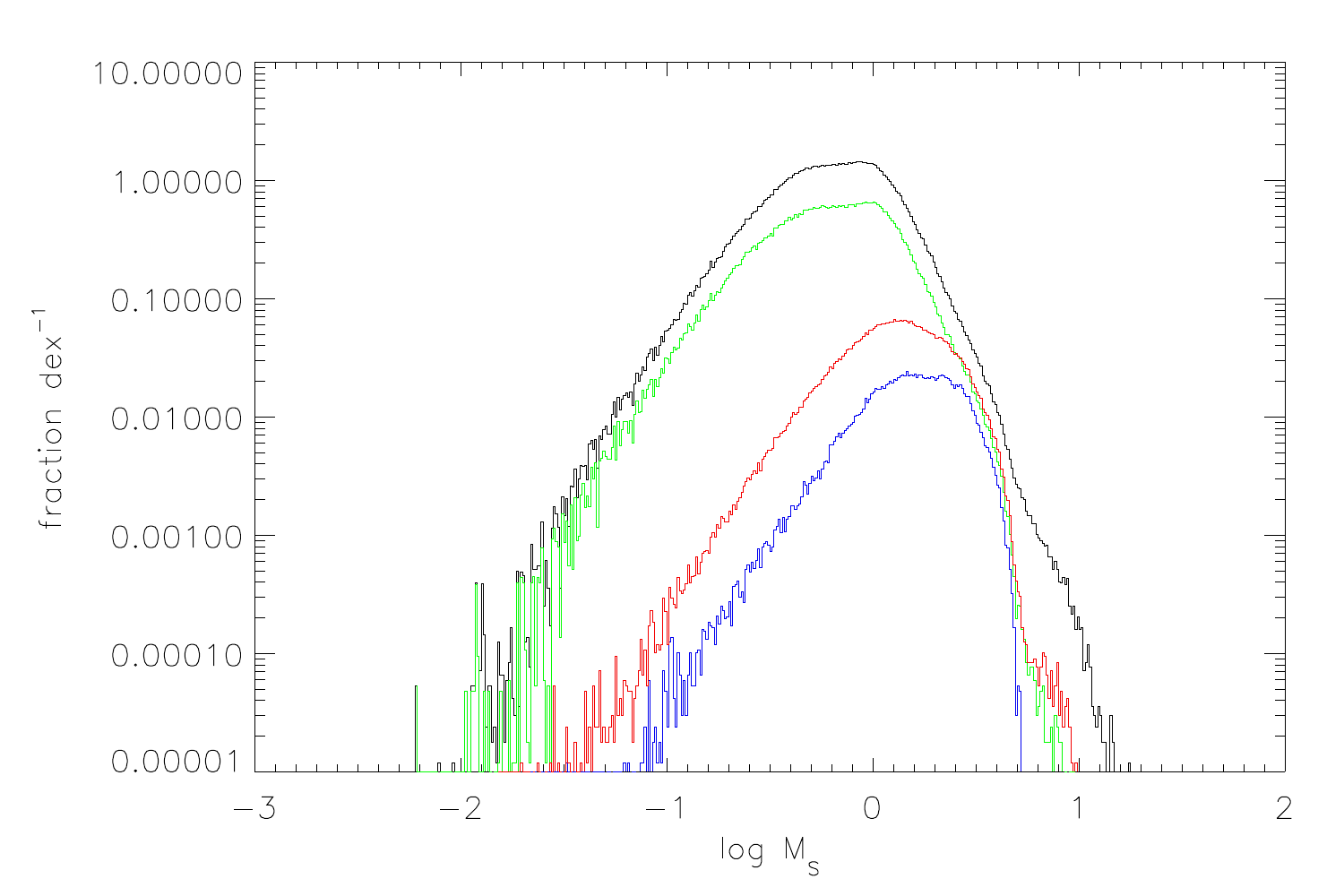}
\caption{Histograms of sonic Mach number for subsets of the hydrodynamical model. {\em Black line}: entire sample. {\em Green line}: all gas at $|z| > 125 \textrm{ pc}$. {\em Red line}: all warm (defined as $5000 < T / \textrm{K} < 20000$) gas. {\em Blue line}: warm gas at $|z| > 125 \textrm{ pc}$. The sound speed in each cell is calculated as $v = 7.18 \, (T/8000 \textrm{ K})^{1/2} \textrm{ km s}^{-1}$, appropriate for a solar metallicity, neutral gas.}
\label{fig:ms_hists}
\end{figure}

\end{document}